\documentstyle[preprint,aps,prb,epsfig]{revtex}
\begin{document}
\draft
\pagestyle{plain}
\newcommand{\D}{\displaystyle}
\newcommand{\ab}{\frac{c_x}{c_y}}
\title{\bf The phase transition phenomena in anisotropic superconductors 
: effect of the orthorhombic crystal field and the potential impurity 
scattering.} 
\author{Grzegorz Hara\'n\cite{AA}, Jason Taylor and A. D. S. Nagi}
\vspace{0.4cm}
\address{Department of Physics, University of Waterloo, 
Waterloo, Ontario, Canada, N2L 3G1}
\date{December 5, 1996} 
\maketitle

\begin{abstract}
A combined effect of the orthorhombic crystal field and potential impurity 
scattering on several superconducting states of a tetragonal symmetry is 
studied within a weak-coupling mean field approach. It is shown that the 
nonmagnetic impurities stabilize the states belonging to the identity irreducible 
representation. The electronic specific heat jump at the phase transition 
is analyzed. Its dependence on the potential scattering rate for large   
impurity concentration is shown  
to be remarkably different for the states with a nonzero value of the 
Fermi surface averaged order parameter than for those with a vanishing one. 
In particular, very distinct signals from $d_{x^2-y^2}$ state in YBCO and 
$d_{xy}$ state in BSCCO compound are predicted. This effect may be 
used as a test for the presence of these states in the above cuprates.\\ 
\end{abstract}

\vspace{0.5cm}
\pacs{Keywords: d-wave superconductor, Fermi surface, pair breaking, 
specific heat at $T_c$, weak coupling} 

\section{Introduction} 
It is generally agreed that a key structural element of the high-$T_c$ 
superconductors is the quasi-two dimensional copper-oxygen plane. 
In most of the cuprates this plane has nearly a tetragonal symmetry with 
a small orthorhombic distortion. \cite{1} 
The unit cells of some of these materials contain other elements,  
like the one-dimensional copper-oxygen chains in ${\rm YBa_2Cu_3O_{7-\delta}}$ 
(YBCO) which 
reduce their symmetry to the orthorhombic one ($C_{2v}$). It is possible 
that the Cu-O chains may also contribute to the superconducting behavior of  
YBCO as the in-plane transport properties display an appreciable 
anisotropy significantly larger than that expected from a slight 
orthorhombicity of the ${\rm CuO_2}$ planes. Several experiments 
\cite{2,3,5,6,7,4,11,8} were 
designed to measure the transport properties separately for the $a$ and $b$  
directions of the copper-oxygen planes. In YBCO, a large anisotropy of 
about $50\%$ between the $a$ and $b$ directions has been observed \cite{2,3} in 
the zero temperature penetration depth ($\lambda_a(0)/\lambda_b(0)\approx 1.55$) 
and a smaller anisotropy has been seen in the thermal conductivity 
(maximum value of $\kappa_a/\kappa_b\approx 1.15$ occurring near 40K). \cite{5,6}  
Also the microwave absorption measurements \cite{3,7} in YBCO 
show a significant in-plane anisotropy in the surface resistance 
and in the infrared conductivity, the later remaining about a factor 
of 2 larger in the $b$ direction for the temperature range of 10-90K.   
The in-plane anisotropy of the surface resistance \cite{3} in the normal state of 
YBCO agrees with the observed dc resistivity anisotropy \cite{4,11}   
$\varrho_a/\varrho_b\sim 2$, which was indicated by the band-structure 
calculations \cite{9} as well. Although very weak orthorhombicity due to 
a superlattice distortion in the Bi-O layer, \cite{1}  
${\rm Bi_2Sr_2CaCu_2O_8}$ (BSCCO) displays $ab$-plane anisotropy \cite{8} 
of about $10\%$ in the optical conductivity and the dc resistivity.    
It has been pointed out, \cite{10,12} that in an orthorhombic system the 
superconducting order parameter becomes a mixture of $d$-wave and $s$-wave  
components as they belong to the same irreducible representation of the crystal 
point group. This leads to an interesting feature in the specific heat jump at 
the phase transition in the presence of a potential impurity scattering. \cite{21}\\  
\indent 
Using a simplified single-band model simulating the orthorhombic anisotropy of the 
Fermi surface for YBCO and BSCCO, we have analyzed a group of superconducting 
states in the untwinned systems. 
We have studied the effect of potential scattering on the stability of these states 
and also on the jump in the electronic specific heat at the phase transition. 

\section{Order parameter and Fermi surface anisotropy}
We consider a singlet superconducting order parameter 
with its orbital part defined as follows

\begin{equation}
\label{e2}
\Delta\left({\bf k}\right)=\Delta e\left({\bf k}\right)
\end{equation}

\noindent
where $e\left({\bf k}\right)$ is a basis real function of a one-dimensional 
(1D) irreducible representation of $C_{4v}$ point group.  
We normalize $e\!\left({\bf k}\right)$ by taking its average value over the 
Fermi surface (FS) $\left<e^{2}\right>=\int_{FS}dS_k n\left({\bf k}\right)
e^2\left({\bf k}\right)=1$, where $\int_{FS}dS_{k}$ represents the integration 
over the Fermi surface and $n\left({\bf k}\right)$ is the angle resolved  
FS density of states, which obeys $\int_{FS}dS_k n\left({\bf k}\right)=1$. 
This normalization gives $\Delta$ the meaning of the absolute magnitude of  
the order parameter. Our discussion is limited to the functions 
$e\left({\bf k}\right)$ confined to the XY plane only, which seems to be    
appropriate for the high $T_{c}$ compounds.\cite{12} These functions  
are listed in Tab. 1, with the symmetry group notation after Ref. 14.  
The basis functions of the $C_{4v}$ irreducible representations $\Gamma_{1}^{+}$, 
$\Gamma_{3}^{+}$ and $\Gamma_{4}^{+}$ are taken as three linearly independent  
second order polynomials. A fourth order polynomial  
is used then as a basis function of the remaining $\Gamma_{2}^{+}$ 
representation. Finally, for the sake of comparison, we consider two more 
fourth order polynomials belonging to the identity irreducible representation  
$\Gamma_{1}^{+}$. The superconducting states given by the functions 
$e\left({\bf k}\right)$ constructed from the angular momentum eigenfunctions  
corresponding to a quantum number L=2 are called the $d$-wave states (second 
order polynomials) and those for which the L=4 eigenfunctions were used 
are called the $g$-wave states (fourth order polynomials). 
Unless an ambiguity arises we will use a more informal name of extended s-wave 
for the states which are invariant under rotation through $\pi/2$ about the $z$  
axis, that is for $\Gamma_{1}^{+}$ and $\Gamma_{2}^{+}$ of $C_{4v}$ irreducible 
representations given in Tab. 1.  

The presence of the orthorhombic anisotropy means that the relevant point  
group is a subgroup of the square, which is 
not the same for all the cuprate superconductors. \cite{12} In the case of YBCO, 
the $a$- and $b$- crystal axes become inequivalent while in BSCCO  
the two orthogonal $45^{\circ}$ axes do so. Therefore the group classification 
of the superconducting states differs in YBCO and BSCCO compounds 
(see Tab. 1). A rotation of a coordinate system through $45^{\circ}$ 
about the z-axis transforms a YBCO-type geometry into a BSCCO-type one and also  
shows an equivalence of the superconducting states, which we summarize in Tab. 2.  
Distinguishing between these two  
symmetries, we make a simple approximation of the orthorhombic  
anisotropy by assuming the following form of the electron band \cite{10,14}   
energy $\xi_{\bf k}$ measured from the Fermi energy $\varepsilon_{F}$ level    

\begin{equation}
\label{e2a}
\xi_{\bf k}=c_xk_{x}^{2}+c_yk_{y}^{2}-\varepsilon_{F},\;\;\;\;\;\;
{\rm for\;\;YBCO} 
\end{equation}

\begin{equation}
\label{e2b}
\xi_{\bf k}=\frac{1}{2}\left(c_x+c_y\right)\left(k_{x}^{2}+k_{y}^{2}\right)
+\left(c_x-c_y\right)k_{x}k_{y}-\varepsilon_{F},\;\;\;\;\;\;
{\rm for\;\;BSCCO} 
\end{equation}

\noindent
A dimensionless ratio of the effective masses $c_x/c_y$ becomes a parameter 
describing the orthorhombic anisotropy of the Fermi surface. It changes  
from 0 to 1, with $c_x/c_y=1$ describing a circular FS and $c_x/c_y=0$ 
corresponding to a one-dimensional limit, non physical for the cuprates.    
For the sake of simplicity the orthorhombic symmetry is introduced as a 
deviation from a cylindrical FS not from a tetragonal one.  
We assume that this simplified single-band model \cite{10} given by Eqs. (\ref{e2a})  
and (\ref{e2b}) captures some essential anisotropic features of both types 
of orthorhombicity i.e. that of YBCO and BSCCO. Particularly in the  
case of YBCO, where the symmetry lowering is mainly due to the existence of 
the chains, limiting the superconducting electrons to the ${\rm CuO_2}$ planes only    
and introducing the crystal field anisotropy by an elliptical Fermi surface 
is to be understood as a first approximation which reflects the basic properties 
related to the orthorhombicity of the compound.    
We employ the above single-band model to study the effect of orthorhombic 
anisotropy on the critical temperature $T_c$ and the specific heat jump at 
the phase transition in the presence of nonmagnetic impurity scattering.  

\section{Nonmagnetic impurity scattering near ${\bf T_c}$}
The single-particle Green's function in the presence of nonmagnetic, 
noninteracting impurities is expressed in Nambu space as  

\begin{equation}
\label{e4}
\hat{G}\left(\omega,{\bf k}\right)=-\frac{1}
{\tilde{\omega}^{2}+{\xi_{k}}^{2}+|\tilde{\Delta}\left(
{\bf k}\right)|^{2}}\left(i\tilde{\omega}\hat{\tau}_0  
+\xi_{k}\hat{\tau}_3+\tilde{\Delta}\left({\bf k}\right)\hat{\tau}_2\right)  
\end{equation}

\noindent
where $\hat{\tau}_0$ and $\hat{\tau}_i$ ($i=1,2,3$) represent the unit and 
Pauli matrices in particle-hole space respectively and $\xi_{k}$ is 
the quasiparticle energy (Eqs. (\ref{e2a}), (\ref{e2b})).  
The renormalized Matsubara frequency $\tilde{\omega}\left({\bf k}\right)$ 
and the renormalized order parameter $\tilde{\Delta}\left({\bf k}\right)$  
are given by 

\begin{eqnarray}
\label{e6}
\tilde{\omega}=\omega-\Sigma_0,&\;\;\;\;&
\tilde{\Delta}\left({\bf k}\right)=\Delta\left({\bf k}\right)+\Sigma_1
\end{eqnarray}

\noindent
with $\omega=\pi T(2n+1)$ (T is the temperature, n is an integer). 
$\Sigma_0$, $\Sigma_1$ 
are the self-energies due to the electron-impurity scattering obtained in 
the t-matrix approximation. \cite{15,16} 
This approach introduces two parameters describing the scattering 
process: $c=1/(\pi N_0 V_i)$ and $\Gamma=n_i/\pi N_0$, where 
$N_{0}$, $V_i$ and $n_i$ are respectively the overall density of  
states at the Fermi surface (FS), the impurity (defect) potential 
and the impurity concentration. We assume $s$-wave scattering by the impurities, 
that is, $V_i$ does not have an internal momentum-dependence. \cite{27}   
It is particularly convenient to think of $c$ as a measure of the  
scattering strength, with $c=0$ in the unitary limit and $c\gg 1$ for 
weak scattering that is the Born limit. 
Assuming a particle-hole symmetry of the quasiparticle spectrum we get 
the self-energies defined as follows

\begin{eqnarray}
\label{e6a}
\D\Sigma_0=-\Gamma\frac{g_0}{c^2+g^{2}_{0}+g^{2}_{1}},&\;\:\;\;&
\D\Sigma_1=\Gamma\frac{g_1}{c^2+g^{2}_{0}+g^{2}_{1}}
\end{eqnarray}

\noindent
with $g_0$, $g_1$ functions determined by the self-consistent equations 

\begin{equation}
\label{e6b}
\D g_0=\frac{1}{N_0\pi}\sum_{\bf k}\frac{\tilde{\omega}}
{\tilde{\omega}^{2}+{\xi_{k}}^{2}+|\tilde{\Delta}\left({\bf k}\right)|^{2}}
\end{equation}

\begin{equation}
\label{e6c}
\D g_1=\frac{1}{N_0\pi}\sum_{\bf k}\frac{\tilde{\Delta}\left({\bf k}\right)} 
{\tilde{\omega}^{2}+{\xi_{k}}^{2}+|\tilde{\Delta}\left({\bf k}\right)|^{2}}
\end{equation}

\noindent
The self-consistency equation for the order parameter reads

\begin{equation}
\label{e7}
\Delta\left({\bf k}\right)=-T\sum_{\omega}\sum_{{\bf k'}}
V\left({\bf k}, {\bf k'}\right)
\frac{\tilde{\Delta}\left({\bf k'}\right)}
{\tilde{\omega}^{2}+{\xi_{k'}}^{2}+|\tilde{\Delta}\left(
{\bf k'}\right)|^{2}}
\end{equation}

\noindent
where $V\left({\bf k}, {\bf k'}\right)$ is the phenomenological pair potential
taken as

\begin{equation}
\label{e8}
V\left({\bf k}, {\bf k'}\right)=-V_{0}e\left({\bf k}\right)
e\left({\bf k'}\right)
\end{equation}

\noindent 
To proceed further, we restrict the wave vectors of the electron self-energy 
and pairing potential to the Fermi surface and replace $\sum_{\bf k}$  
by $N_{0}\int_{FS}dS_{k}n\left({\bf k}\right)\int d\xi_{k}$. 
Integrating over $\xi_{k}$, the gap equation (\ref{e7}) can be transformed 
after a standard procedure \cite{17} into 

\begin{equation}
\label{e8a}
\ln\left(\frac{T}{T_{c_{0}}}\right)=2\pi T\sum_{\omega\ge 0}
\left(\int_{FS}dS_{k}n\left({\bf k}\right) 
f\left(\omega,{\bf k}\right)-\frac{1}{\omega}\right)
\end{equation}

\noindent
where the $f\left(\omega,{\bf k}\right)$ function is defined as follows

\begin{equation}
\label{e8b}
\D f\left(\omega,{\bf k}\right)=  
\frac{e\left({\bf k}\right)\tilde{\Delta}\left({\bf k}\right)}
{\Delta\left[\tilde{\omega}^{2}+|\tilde{\Delta}\left(
{\bf k}\right)|^{2}\right]^{\frac{1}{2}}}
\end{equation}

\noindent
and $T_{c_{0}}$ is the critical temperature in the absence of impurities. 
We expand the gap equation (\ref{e8a}) in powers of $\Delta^{2}$ around
$\Delta=0$ taking into account that $\tilde{\omega}$ and $\tilde{\Delta}$ 
are functions of $\Delta^{2}$ as given by Eqs. (\ref{e6})-(\ref{e6c}). 
Keeping up to the second power terms in $\Delta$ we get
the Ginzburg-Landau approximation of the gap equation 

\begin{equation}
\label{e9}
\ln\left(\frac{T}{T_{c_{0}}}\right)=-f_{0}-\frac{1}{2}f_{1}
\left(\frac{\Delta}{2\pi T}\right)^{2}
\end{equation}

\noindent
where the coefficients are given by

\begin{equation}
\label{e10}
f_{0}=-2\pi T\sum_{\omega>0}\left(\int_{FS}dS_{k}n\left({\bf k}\right)
\left(f\left(\omega,{\bf k}\right)\right)_{\Delta=0}
-\frac{1}{\omega}\right)
\end{equation}

\begin{equation}
\label{e11}
f_{1}=-\left(2\pi T\right)^{3}\sum_{\omega}
\int_{FS}dS_{k}n\left({\bf k}\right) 
\left(\frac{df\left(\omega,{\bf k}\right)}{d\Delta^{2}}\right)_{\Delta=0}
\end{equation}

\noindent
Taking the derivatives with respect to $\Delta^2$ 

\begin{equation}
\label{e25}
\D\frac{d\;\;\;}{d\Delta^{2}}= \frac{\partial\;\;\;}{\partial\Delta^{2}}+
\sum_{\omega}\left\{\frac{d\tilde{\omega}}
{d\Delta^{2}}\frac{\partial\;\;}{\partial\tilde{\omega}}  
+\frac{d\tilde{\Delta}\left({\bf k}\right)}
{d\Delta^{2}}\frac{\partial\;\;\;\;\;\;}
{\partial\tilde{\Delta}\left({\bf k}\right)}\right\}   
\end{equation}

\noindent 
and with a use of the relations given in Eqs. (\ref{e6})-(\ref{e6c}) 
and (\ref{e8b}) we obtain 

\begin{equation}
\label{e13}
\begin{array}{l}
\D f_{0}=\left(1-\left<e\right>^{2}\right)
\left(\psi\left(\frac{1}{2}+\varrho\right)
-\psi\left(\frac{1}{2}\right)\right)\;\;\;\;\;\;\;\;\;\;\;\;\;\;\; 
\;\;\;\;\;\;\;\;\;\;\;\;\;\;\;\;\;\;\;\;\;\;\;\;\;\;\;\;\;\;\;\;\;
\;\;\;\;\;\;\;\;\;\;\;\;\;
\end{array}
\end{equation}

\vspace{2ex}  

\begin{equation}
\label{e14a}
\begin{array}{l}
\D f_{1}=2\left<e\right>\left[2\left<e^3\right>+5\left<e\right>^3 
-7\left<e\right>\right]\varrho^{-2}\left(\psi\left(\frac{1}{2}
+\varrho\right)-\psi\left(\frac{1}{2}\right)\right)\\
\\
\D +2\left<e\right>\left[-2\left<e^{3}\right>-3\left<e\right>^3 
+5\left<e\right>\right] 
\varrho^{-1}\psi^{(1)}\left(\frac{1}{2}+\varrho\right)
+4\left<e\right>^2\left[1-\left<e\right>^{2}\right]\varrho^{-1} 
\psi^{(1)}\left(\frac{1}{2}\right)\\
\\
\D +\frac{1}{2}\left[-\left<e^{4}\right>+3\left<e\right>^{4}
+4\left<e\right>\left<e^{3}\right>-6\left<e\right>^{2}\right] 
\psi^{(2)}\left(\frac{1}{2}+\varrho\right)
-\frac{1}{2}\left<e\right>^{4}\psi^{(2)}\left(\frac{1}{2}\right)\\
\\
\D +\frac{1}{6}\left[2\left(\left<e\right>^{2}-1\right)^2\frac{1}{c^2+1}
-\left<e\right>^{4}+2\left<e\right>^{2}-1\right] 
\varrho\psi^{(3)}\left(\frac{1}{2}+\varrho\right)
\end{array}
\end{equation}

\noindent
where $\varrho=\Gamma/\left[(c^2+1)2\pi T\right]$ and $\psi$, $\psi^{(n)} 
\;(n=1,2,3)$ are the polygamma functions. \cite{19}  
In the unitary limit $c=0$ and $\varrho=\Gamma/(2\pi T)$. Alternatively for 
weak scattering ($c\gg 1$) we obtain the Born scattering   
rate $\varrho=\pi N_0 n_i V^2_i/(2\pi T)$ and also neglect  
$\varrho/(c^2+1)$ in the last term of Eq. (\ref{e14a}).   
Coefficients $f_0$ and $f_1$ involve three different types of  
the Fermi surface averages of the superconducting order parameter namely, 
$\left<e\right>$, $\left<e^{3}\right>$, and $\left<e^{4}\right>$, which  
depend on the orthorhombic anisotropy parameter $c_x/c_y$. \cite{28} These 
averages enter the free energy and determine the thermodynamic properties 
at the phase transition. While the critical temperature $T_c$ is determined  
by the $f_0$ function and therefore is characterized only by $\left<e\right>$, 
the other thermodynamic quantities like the entropy or the specific heat 
for instance, involve, through the $f_1$ function (Eq. (\ref{e14a})), the Fermi 
surface average values of higher powers of $e\left({\bf k}\right)$.  
It is illustrative for the purpose of this paper to show these averages as  
functions of the orthorhombic anisotropy parameter $c_x/c_y$.  
We present $\left<e\right>$, $\left<e^{3}\right>$, and 
$\left<e^{4}\right>$ in Figs. 1a - 1c respectively, where the curve numbers 
correspond to the states listed in Tab. 2. The analytical expressions of 
these averages for the YBCO symmetry (Eq. (\ref{e2a}))  are given in the 
Appendix. It is worth mentioning here, that 
as the $k^2_x-k^2_y$ state ($d_{x^2-y^2}$) belongs to different irreducible  
representations (Tab. 1) in YBCO ($\Gamma^{+}_1$) and BSCCO ($\Gamma^{+}_3$),  
its FS average value $\left<e\right>$ is non zero in the first compound 
($c_x/c_y<1$), but $\left<e\right>$ is unchanged by the orthorhombic 
crystal field and is zero in the second one. Further, for the $k_xk_y$ 
state ($d_{xy}$) $\left<e\right>=0$ in YBCO ($\Gamma^{+}_3$) and 
$\left<e\right>\neq 0$ in BSCCO ($\Gamma^{+}_1$). These facts are of crucial  
importance for the critical temperature and the jump in the electronic  
specific heat at the phase transition, which we discuss in the following 
sections.    

\section{Critical temperature ${\bf T_c}$}
We analyze in the Ginzburg-Landau (G-L) regime the stability  
of different superconducting states in the presence of impurity potential 
scattering and the orthorhombic crystal field. For any FS we consider the states 
of the same critical temperature in the absence of impurities ($T_{c_{0}}$), 
which means that we normalize the order parameter eigenfunctions as  
$\left<e^2\right>=1$. \cite{29} Then we look at the impurity effect on $T_c$ of 
these states. It should be noted that since the orthorhombic anisotropy changes   
the density of states at the Fermi level, $T_{c_{0}}$ is different for different 
values of $c_x/c_y$ even with the condition $\left<e^2\right>=1$.  
We are interested in the influence of impurity scattering on the states of different   
symmetry but with the same $T_{c_{0}}$ in the presence of a given orthorhombic 
crystal field i.e. for a given $c_x/c_y$ value.   
The G-L free energy difference $\Delta F=F_s-F_n$ between the superconducting 
($F_s$) and the normal ($F_n$) phase written up to the fourth order terms  
in $\Delta$ 

\begin{equation}
\label{e14c} 
\Delta F=\alpha\Delta^2+\frac{1}{2}\beta\Delta^4
\end{equation}

\noindent
with the coefficients $\alpha$ and $\beta$ determined from Eqs. (\ref{e9}), 
(\ref{e13}) and (\ref{e14a}) i.e.  

\begin{equation}
\label{e14d}
\alpha=N_0 \left(\ln\left(\frac{T}{T_{c_{0}}}\right)+f_0\right) 
\end{equation}

\begin{equation}
\label{e14e}
\beta=\frac{N_0 f_1}{2\left(2\pi T_c\right)^2} 
\end{equation}

\noindent
leads to the free energy minimum given by 

\begin{equation}
\label{e14f}
\left(\Delta F\right)_{min}=-\frac{1}{2}\frac{\alpha^2}{\beta}
\end{equation}

\noindent
One can see from Eqs. (\ref{e13}) and (\ref{e14d}) that when $\alpha=0$ for  
a state with $\left<e\right>=0$ then its value is less than zero for a nonzero 
$\left<e\right>$. Therefore according to Eq. (\ref{e14f})   
even an infinitesimal impurity scattering rate stabilizes states with 
$\left<e\right>\neq 0$ over $\left<e\right>=0$ states. In other words 
the critical temperature in the presence of nonmagnetic impurities given 
by \cite{22,23}  

\begin{equation}
\label{e14b}
\ln\left(\frac{T_c}{T_{c_{0}}}\right)=\left(\left<e\right>^{2}-1\right)
\left(\psi\left(\frac{1}{2}+\varrho_c\right)
-\psi\left(\frac{1}{2}\right)\right)
\end{equation} 

\noindent
with $\varrho_c$ as the value of $\varrho$ at $T_c$, 
is higher for the states with a nonzero $\left<e\right>$ value than those  
characterized by $\left<e\right>=0$. It is also worth mentioning that 
the states characterized by the same value of $\left<e\right>$ are degenerate. 
We show the solutions of Eq. (\ref{e14b}) 
for the states from Tab. 2 in the case of a not broken tetragonal symmetry   
($c_x/c_y=1$) in Fig. 2, and for an orthorhombic symmetry 
with $c_x/c_y=0.8$ in Fig. 3. Comparison of Fig. 1a (at $c_x/c_y=1$) and Fig. 2 
shows a degeneracy of the states with the same $\left<e\right>$ value. 
When the FS symmetry is lowered   
to $C_{2v}$ the states fall into two irreducible representations $\Gamma^{+}_1$ 
and $\Gamma^{+}_3$ (Tab. 1). The two $\Gamma^{+}_3$ states (curves 3 and 6  
in Fig. 3) are still degenerate 
as $\left<e\right>=0$ in that case (Fig. 1a). On the other hand most   
of $\Gamma^{+}_1$ states are influenced by the orthorhombicity of the system and  
their critical temperatures are split. We present the evaluation of $T_c$ for the  
states under consideration at the impurity scattering level \cite{26}   
$\varrho_c T_c/T_{c_{0}}=0.1$ as a function of 
orthorhombic anisotropy parameter $c_x/c_y$ in Fig. 4. As mentioned earlier, 
the critical temperature depends on the order parameter and FS symmetries   
through $\left<e\right>$ only. This leads to a remarkable similarity between the   
effect of FS orthorhombicity on $T_c$ (Fig. 4) and on $\left<e\right>$ (Fig. 1a).   

\section{Specific heat jump at the phase transition} 
The specific heat jump at $T_c$,  
$\Delta C(T_c)$, normalized by the normal state specific heat $C_N(T_c)$  
is given by    

\begin{equation}
\label{e14}
\D\frac{\Delta C(T_c)}{C_N(T_c)}=\frac{12}{\left(f_1\right)_{T=T_c}} 
\left[1+T_c\left(\frac{df_0}{dT}\right)_{T=T_c}\right]^{2} 
\end{equation}

\noindent
and $f_0$ from Eq. (\ref{e13}) yields  

\begin{equation}
\label{e15}
\D\frac{\Delta C(T_c)}{C_{N}(T_c)}=\frac{12}{f_{1}\left(\varrho_{c}\right)}
\left[1+\left(\left<e\right>^2-1\right)\varrho_{c}\psi^{(1)}\left(\frac{1}{2}+
\varrho_{c}\right)\right]^{2}
\end{equation}

\noindent
where $\varrho_c$ is $\varrho$ at $T=T_c$. 
This rather cumbersome formula, when considered along with Eq. (\ref{e14a}),  
reduces significantly for $\left<e\right>=0$ case 

\begin{equation}
\label{e15a}
\D\frac{\Delta C(T_c)}{C_{N}(T_c)}=\frac{\D 12\left[1-\varrho_c\psi^{(1)}
\left(\frac{1}{2}+\varrho_c\right)\right]^{2}}
{\D \frac{\mu}{6}\varrho\psi^{(3)}\left(\frac{1}{2}+\varrho_c\right)
-\frac{1}{2}\left<e^{4}\right>
\psi^{(2)}\left(\frac{1}{2}+\varrho_c\right)}
\end{equation}
  
\noindent
where $\mu=(1-c^2)/(1+c^2)$. For an appropriate choice of $\left<e^{4}\right>$  
value $\Delta C(T_c)/C_{N}(T_c)$ from Eq. (\ref{e15a}) agrees  
with the result obtained by Hirschfeld et al. \cite{16} as well as with that 
obtained by Suzumura and Schulz \cite{20} in the Born limit.  
For a pure system \cite{21} where $\varrho_c=0$   

\begin{equation}
\label{e16}
\D\left(\frac{\Delta C\left(T_c\right)}{C_{N}\left(T_c\right)}\right)_{\varrho_c=0}
=-\frac{24}{\D\psi^{(2)}\left(\frac{1}{2}\right)\left<e^{4}\right>}
\approx\frac{1.426}{\left<e^{4}\right>}
\end{equation}

\noindent
Alternatively, in a highly impure superconductor 
with $\varrho_c\rightarrow\infty$ i.e. $T_c\rightarrow 0$ due to   
suppression by the impurities, the two cases, depending on  
$\left<e\right>$ value, are to be distinguished. \cite{21} First, when 
$\left<e\right>\neq 0$ leads to 
  
\begin{equation}
\label{e17}
\D\left(\frac{\Delta C\left(T_c\right)}{C_{N}\left(T_c\right)}\right)
_{\varrho_c\rightarrow\infty}=
-\frac{24}{\D\psi^{(2)}\left(\frac{1}{2}\right)}\approx 1.426 
\end{equation}

\noindent
and the second, with $\left<e\right>=0$ yields   

\begin{equation}
\label{e18}
\D\left(\frac{\Delta C\left(T_c\right)}{C_{N}\left(T_c\right)}\right)
_{\varrho_c\rightarrow\infty}=0
\end{equation}
 
\noindent
We note, that the specific heat jump value in $\varrho_c\rightarrow\infty$ 
limit for a nonzero value of $\left<e\right>$ 
given by Eq. (\ref{e17}) agrees 
with that of an isotropic $s$-wave superconductor. This fact has a simple 
intuitive interpretation. A nonzero Fermi surface average of the 
order parameter leads to an asymptotic power-law critical temperature 
suppression for large impurity concentration \cite{14} 
$T_c\sim\left(T_{c_{0}}\right)^{1/\left<e\right>^2}\left[\Gamma/
\left(c^2+1\right)\right]^{(1-1/\left<e\right>^2)}$, 
therefore $T_c$ is almost constant for 
large $\Gamma$ values. The impurity effect, then, 
in the high impurity concentration range is the same as in the case 
of $s$-wave superconductivity, where $T_c$ is not changed by the 
nonmagnetic impurities. Alternatively, for $\left<e\right>=0$ we observe 
a strong impurity-induced suppression of the critical temperature 
\cite{22,23} leading to its   
zero value at finite impurity concentration, which is reflected by  
a zero specific heat jump limit value in Eq. (\ref{e18}).  
Therefore the jump in the specific heat at $T_c\rightarrow 0$ approaches 
a value of 1.426 for the states belonging to $\Gamma^{+}_1$ representation 
if their $\left<e\right>$ is nonzero, whereas it decreases to zero for 
all the other states described by the non identity  irreducible representations 
for which $\left<e\right>=0$. Indeed, as it has been shown for the representative order 
parameters, \cite{24,25} the gap anisotropy is smeared out by the isotropic 
impurity scattering in the first case ($\left<e\right>\neq 0$) and the 
density of states approaches that of an isotropic $s$-wave superconductor.  
The nonmagnetic impurities,    
however,  are severe pair-breakers in a state with $\left<e\right>=0$ 
and lead to a finite density of states at the Fermi energy. This qualitative 
difference is expected to be reflected in the quantities proportional to the 
density of states, for instance in the specific heat. 
According to the classification given in Tab. 2,    
in the limit of $T_c\rightarrow 0$ one should observe the BCS specific heat jump 
value for a $d_{x^2-y^2}$ state in YBCO ($\Gamma^{+}_1$ representation), 
but the jump should vanish in BSCCO ($\Gamma^{+}_3$ representation). On the 
other hand $d_{xy}$ state in YBCO should lead to a zero jump in $T_c\rightarrow 0$ 
limit, but the BCS-like jump in BSCCO. In the extended $s$-wave state    
$k_xk_y(k_x^2-k_y^2)$, which belongs to $\Gamma^{+}_3$ representation in both 
YBCO and BSCCO structures, the specific heat jump should decrease to zero with   
$T_c\rightarrow 0$ in both compounds. The results for specific heat jump  
in Born and unitary scattering limits are  
shown in Figs. 5 and 6 respectively for a circular Fermi surface ($c_x/c_y=1$).  
The YBCO-type FS with a small orthorhombic anisotropy ($c_x/c_y=0.8$) leads to the 
solutions presented in Fig. 7 (Born scattering) and Fig. 8 (unitary scattering).   
Finally, Figs. 9 and 10 correspond to both considered impurity scattering limits 
in a system with a Fermi surface of a large orthorhombicity given by $c_x/c_y=0.2$.  
In the case of a not broken tetragonal 
symmetry (Figs. 5 and 6) the specific heat jump of the identity representations 
$\Gamma^{+}_1$ with $\left<e\right>\neq 0$ (curves 1,4,5), is almost constant and 
nonzero, but that of the other $\Gamma^{+}_2$, $\Gamma^{+}_3$ and $\Gamma^{+}_4$ 
irreducible representations with $\left<e\right>=0$ (curves 6,2 and 3 
respectively) is suppressed to zero by the impurities.   
It is also remarkable in Figs. 5 and 6 that the states having the same values 
of $\left<e\right>$, $\left<e^3\right>$ and $\left<e^4\right>$ averages   
(compare Figs. 1a-1c at $c_x/c_y=1$) display the same values of the jump in the  
specific heat. When the FS symmetry changes into the orthorhombic there are 
four $\Gamma^{+}_1$ representations in the set of considered functions (Tabs. 1, 2). 
We look at YBCO symmetry first. 
Except for two states of $\Gamma^{+}_3$ symmetry ($k_xk_y$, $k_xk_y(k^2_x-k^2_y)$) 
which are strongly suppressed by the nonmagnetic impurity scattering, the specific 
heat jump of all the other states goes to a BCS value of 1.426 in $T_c\rightarrow 0$ 
limit. For a small orthorhombic anisotropy a dramatic rise of the specific   
heat jump is seen only in the $d_{x^2-y^2}$ state (Figs. 7 and 8),  
whereas for a large FS anisotropy this increase, however not so abrupt, 
is observed for all four identity representations (Figs. 9 and 10).  
In the case of BSCCO symmetry (see Tab. 2) the sharp rise in the specific 
heat jump for a small orthorhombic FS anisotropy (Figs. 7 and 8)  
will be observed only for the $d_{xy}$ state. The $d_{x^2-y^2}$ state will  
not mix with the identity representations in BSCCO, which will lead to   
a zero value of the specific heat jump in the large impurity scattering 
limit. Therefore observation of a sharp rise in the specific heat jump  
at the phase transition with the transition temperature $T_c\rightarrow 0$  
in dirty BSCCO would provide information about the possible presence of $d_{xy}$ state  
in the condensate. Note that in impure YBCO the same observation  
would signal a realization of the $d_{x^2-y^2}$ state. We do not consider the  
mixed states since it would require an analysis of many cases 
even if we confine ourselves to the representations from Tab. 1. 
However, one should take into account that for a small FS anisotropy 
(Figs. 7 and 8) the rise in the specific heat jump  
at $T_c\rightarrow 0$ will be less pronounced for a superposition of some  
$\Gamma^{+}_1$ states than for a single $d_{x^2-y^2}$ state in the case of YBCO or 
a $d_{xy}$ one in BSCCO. Comparing the effect of Born scattering 
on the jump in the specific heat (Figs. 5, 7, 9) with that of the unitary impurity 
scattering (Figs. 6, 8, 10), we notice a difference in the range of medium impurity 
concentration for the states with the small $\left<e\right>$ values (see Fig. 1a). 
It means that the states suppressed by the impurity scattering the most, display 
the largest difference between these two scattering limits, which is particularly 
seen in the states of $\left<e\right>=0$, where the pair-breaking effect of the 
impurities is the strongest (Eq. (\ref{e14b})).   

\section{Conclusion}
The orthorhombic anisotropy of the cuprates results in a number of states belonging 
to the identity irreducible representation ($\Gamma^{+}_1$). For these states 
the average value of the order parameter over the Fermi surface ($\left<e\right>$)    
may not vanish. The behavior for states with $\left<e\right>\neq 0$ is 
qualitatively different than for states with $\left<e\right>=0$.  
We have studied the effect of nonmagnetic impurity scattering on the superconducting   
states which may be realized in the high temperature superconductors, distinguishing 
between the orthorhombicity of YBCO and BSCCO compounds. It has been pointed out, that 
the potential scattering stabilizes the $\Gamma^{+}_1$ states (with $\left<e\right>\neq 0$) 
against those with $\left<e\right>=0$. A very interesting feature may be observed in    
the electronic specific heat jump at the phase transition for a small crystal 
orthorhombicity. The $d_{x^2-y^2}$ state in YBCO and the $d_{xy}$ state in BSCCO  
are predicted to lead to a sharp rise in the jump in specific heat at $T_c$ in  
the limit $T_c\rightarrow 0$ when the critical temperature 
is suppressed by the potential impurity scattering.  
This effect may be helpful in the identification of the superconducting states in 
the cuprates.    

\medskip
\section*{Acknowledgment}
\noindent 
Work supported by the Natural Sciences and Engineering 
Research Council of Canada.

\newpage
\appendix   
\section*{} 
In this appendix we present the Fermi surface averages of the 1-st, 3-rd  
and 4-th powers of the normalized  
order parameter function $e\left({\bf k}\right)$   
for the states considered in the paper. The average values were taken 
over the YBCO-type Fermi surface given by Eq. (\ref{e2a}). To obtain the 
results for BSCCO one should use the equivalence between the states 
in YBCO and BSCCO geometry summarized in Tab. 2.\\  

\begin{equation} 
\underline{e\left({\bf k}\right)=\left(k^{2}_{x}\pm k^{2}_{y}\right)
\left<\left(k^{2}_{x}\pm k^{2}_{y}\right)^{2}\right>^{-1/2}}
\;\;\;\;\;\;\;\;\;\;\;\;\;\;\;\;\;\;\;\;\;\;\;\;\;\;\;\;\;\;
\;\;\;\;\;\;\;\;\;\;\;\;\;\;\;\;\;\;\;\;
\:\;\;\;\;\;\;\;\;\;\;\;\;\;\;\;\;\;\;\;\;\;
\end{equation} 

\begin{displaymath} 
\begin{array}{l} 
\D\nu\left(\ab\right)=\left[\frac{3}{2}\pm\ab
+\frac{3}{2}\left(\ab\right)^{2}\right]^{-1/2}\\ 
\\
\D\left<e\right>=\nu\left(\ab\right)
\left[1\pm\ab\right]\\ 
\\
\D\left<e^{3}\right>=\nu^{3}\left(\ab\right)
\left[\frac{5}{2}\left(1\pm\left(\ab\right)^{3}\right)
\pm\frac{3}{2}\ab\left(1\pm\ab\right)\right]\\
\\
\D\left<e^{4}\right>=16\nu^{4}\left(\ab\right)
\left[\frac{35}{128}\left(1\mp\ab\right)^{4}
-\frac{5}{4}\left(1\mp\ab\right)^{3}
+\frac{9}{4}\left(1\mp\ab\right)^{2}
-2\left(1\mp\ab\right)+1\right]
\end{array}
\end{displaymath}

\vspace{0.5cm}

\begin{equation}
\underline{e\left({\bf k}\right)=\left(k_xk_y\right)
\left<\left(k_xk_y\right)^{2}\right>^{-1/2}}
\;\;\;\;\;\;\;\;\;\;\;\;\;\;\;\;\;\;\;\;\;\;\;\;\;\;\;\;\;\;
\;\;\;\;\;\;\;\;\;\;\;\;\;\;\;\;\;\;\;\;\;\;
\:\;\;\;\;\;\;\;\;\;\;\;\;\;\;\;\;\;\;\;\;\;\;\;
\end{equation}

\begin{displaymath}
\begin{array}{l} 
\left<e\right>=0\;,\;\;\;\left<e^{3}\right>=0\;,\;\;\;\left<e^{4}\right>=
\frac{3}{2}\;\;\;\;\;\;\;\;\;\;\;\;\;\;\;\;\;\;\;\;\;\;\;\;\;\;\;\;\;\;
\;\;\;\;\;\;\;\;\;\;\;\;\;\;\;\;\;\;\;\;\;\;\;\;
\:\;\;\;\;\;\;\;\;\;\;\;\;\;\;\;\;\;\;\;\;\;\;\;
\end{array}
\end{displaymath}

\vspace{0.5cm}

\begin{equation}
\underline{e\left({\bf k}\right)=\left(k^2_xk^2_y\right)
\left<\left(k^2_xk^2_y\right)^{2}\right>^{-1/2}}
\;\;\;\;\;\;\;\;\;\;\;\;\;\;\;\;\;\;\;\;\;\;\;\;\;\;\;\;\;\;
\;\;\;\;\;\;\;\;\;\;\;\;\;\;\;\;\;\;\;\;
\:\;\;\;\;\;\;\;\;\;\;\;\;\;\;\;\;\;\;\;\;\;\;\;
\end{equation}

\begin{displaymath}
\begin{array}{l} 
\D\left<e\right>=\left(\frac{2}{3}\right)^{1/2}\;,\;\;\; 
\left<e^{3}\right>=\frac{5}{3}\left(\frac{2}{3}\right)^{1/2}\;,\;\;\; 
\left<e^{4}\right>=\frac{35}{18}\;\;\;\;\;\;\;\;\;\;\;\;\;\;\;\;\;\;\;\;
\;\;\;\;\;\;\;\;\;\;\;\;\;\;\;\;\;\;\;\;\;\;\;\;\;\;\;\;\;\;\;
\end{array}
\end{displaymath}

\vspace{0.5cm}

\begin{equation}
\underline{e\left({\bf k}\right)=\left(k_xk_y\left(k^2_x-k^2_y\right)\right)
\left<\left(k_xk_y\left(k^2_x-k^2_y\right)\right)^{2}\right>^{-1/2}}
\;\;\;\;\;\;\;\;\;\;\;\;\;\;\;\;\;\;\;\;
\;\;\;\;\;\;\;\;\;\;\;\;\;\;\;\;\;\;\;\;
\;\;\;\;\;\;\;\;\;\;\;\;\;\;\;\;\;\;\;\;
\end{equation}

\begin{displaymath}
\begin{array}{l}
\left<e\right>=0\;,\;\;\;\left<e^{3}\right>=0\\
\\
\D\left<e^{4}\right>=\left[\frac{5}{128}\left(1+\left(\ab\right)^{2}\right)
-\frac{3}{64}\ab\right]^{-2}
\left[d_{1}\left(1+\left(\ab\right)^{4}\right)
-4d_{2}\ab\left(1+\left(\ab\right)^{2}\right)
+6d_{3}\left(\ab\right)^{2}\right]
\end{array}
\end{displaymath}

\vspace{0.5cm}

\begin{equation}
\underline{e\left({\bf k}\right)=\left(k^2_x-k^2_y\right)^2 
\left<\left(k^2_x-k^2_y\right)^{4}\right>^{-1/2}}
\;\;\;\;\;\;\;\;\;\;\;\;\;\;\;\;\;\;\;\;\;\;\;\;\;\;\;\;\;\;
\;\;\;\;\;\;\;\;\;\;\;\;\;\;\;\;\;\;\;\;
\:\;\;\;\;\;\;\;\;\;\;\;\;\;\;\;\;\;\;\;
\end{equation}

\begin{displaymath}
\begin{array}{l} 
\D\nu\left(\ab\right)= 
\left[\frac{35}{128}\left(1+\ab\right)^{4}-\frac{5}{4}\left(1+\ab\right)
^{3}+\frac{9}{4}\left(1+\ab\right)^{2}-2\left(1+\ab\right)+1\right]^{-1/2}\\
\\
\D\left<e\right>=\nu\left(\ab\right)
\frac{1}{4}\left(\frac{3}{2}-\ab+\frac{3}{2}\left(\ab\right)^{2}\right)\\ 
\\
\D\left<e^{3}\right>=\nu^3\left(\ab\right)
\left[d_{6}\left(1+\left(\ab\right)^{6}\right)-6d_{7}\ab  
\left(1+\left(\ab\right)^{4}\right)+15d_{8}\left(\ab\right)^{2}
\left(1+\left(\ab\right)^{2}\right)\right.\\
\\
\D\left.-20d_{9}\left(\ab\right)^{3}\right]\\  
\\
\D\left<e^{4}\right>=\nu^4\left(\ab\right)
\left[d_{4}\left(1+\left(\ab\right)^{8}\right)-8d_{5}\ab
\left(1+\left(\ab\right)^{6}\right)+28d_{1}\left(\ab\right)^{2}
\left(1+\left(\ab\right)^{4}\right)\right.\\
\\
\D\left.-56d_{2}\left(\ab\right)^{3}\left(1+\left(\ab\right)^{2}\right)
+70d_{3}\left(\ab\right)^{4}\right]
\end{array}
\end{displaymath}
\noindent
\vspace{0.2cm}\\
where
\vspace{0.2cm}\\

\begin{displaymath}
\begin{array}{lllll}
\D d_{1}=\frac{99}{32768}\;,& \D d_{2}=\frac{45}{32768}\;,&
\D d_{3}=\frac{35}{32768}\;,& \D d_{4}=\frac{6435}{32768}\;,& 
\D d_{5}=\frac{429}{32768}\;,\\
\\
\D d_{6}=\frac{231}{1024}\;,& \D d_{7}=\frac{21}{1024}\;,& 
\D d_{8}=\frac{7}{1024}\;,& \D d_{9}=\frac{5}{1024}\;.&
\end{array}
\end{displaymath}

\newpage
\thispagestyle{empty}
\begin{center} 
{\large FIGURE CAPTIONS}
\end{center} 
\noindent
Fig. 1. The orthorhombic crystal field $c_x/c_y$ dependence of the Fermi 
surface averaged powers of the superconducting order parameter 
function $e\left({\bf k}\right)$: a) $\left<e\right>$, b) $\left<e^3\right>$, 
c) $\left<e^4\right>$. The figure labels correspond to the state numbers 
in Tab. 2.\\  

\noindent
Fig. 2. Normalized critical temperature $T_c/T_{c_{0}}$ of the states 
from Tab. 2 as a function of the normalized impurity scattering rate   
$\varrho_c T_c/T_{c_{0}}$ for the circular Fermi surface ($c_x/c_y=1$).\\  

\noindent
Fig. 3. Normalized critical temperature $T_c/T_{c_{0}}$ of the states 
from Tab. 2 as a function of the normalized impurity scattering rate  
$\varrho_c T_c/T_{c_{0}}$ for the elliptical Fermi surface ($c_x/c_y=0.8$).\\  

\noindent
Fig. 4. Normalized critical temperature $T_c/T_{c_{0}}$ of the states
from Tab. 2 as a function of the Fermi surface orthorhombic anisotropy 
parameter $c_x/c_y$ for the impurity scattering rate 
$\varrho_c T_c/T_{c_{0}}=0.1$.\\   

\noindent
Fig. 5. Jump in the specific heat at $T_c$ normalized by the normal specific  
heat at $T_c$ of the states from Tab. 2 as a function of the normalized 
impurity scattering rate $\varrho_c T_c/T_{c_{0}}$ in the Born limit  
for the circular Fermi surface ($c_x/c_y=1$).\\  

\noindent
Fig. 6. Jump in the specific heat at $T_c$ normalized by the normal specific
heat at $T_c$ of the states from Tab. 2 as a function of the normalized 
impurity scattering rate $\varrho_c T_c/T_{c_{0}}$ in the unitary limit 
for the circular Fermi surface ($c_x/c_y=1$).\\

\noindent
Fig. 7. Jump in the specific heat at $T_c$ normalized by the normal specific
heat at $T_c$ of the states from Tab. 2 as a function of the normalized 
impurity scattering rate $\varrho_c T_c/T_{c_{0}}$ in the Born limit   
for the elliptical Fermi surface ($c_x/c_y=0.8$).

\newpage 
\thispagestyle{empty}
\noindent
Fig. 8. Jump in the specific heat at $T_c$ normalized by the normal specific
heat at $T_c$ of the states from Tab. 2 as a function of the normalized 
impurity scattering rate $\varrho_c T_c/T_{c_{0}}$ in the unitary limit 
for the elliptical Fermi surface ($c_x/c_y=0.8$).\\

\noindent
Fig. 9. Jump in the specific heat at $T_c$ normalized by the normal specific
heat at $T_c$ of the states from Tab. 2 as a function of the normalized 
impurity scattering rate $\varrho_c T_c/T_{c_{0}}$ in the Born limit 
for the elliptical Fermi surface ($c_x/c_y=0.2$).\\

\noindent
Fig. 10. Jump in the specific heat at $T_c$ normalized by the normal specific
heat at $T_c$ of the states from Tab. 2 as a function of the normalized 
impurity scattering rate $\varrho_c T_c/T_{c_{0}}$ in the unitary limit 
for the elliptical Fermi surface ($c_x/c_y=0.2$). 

\newpage
\thispagestyle{empty}
  \parbox{1.5cm}{\vfill $$ \left<e\right> $$ \vfill }
  \parbox{13.5cm}{\epsfig{file=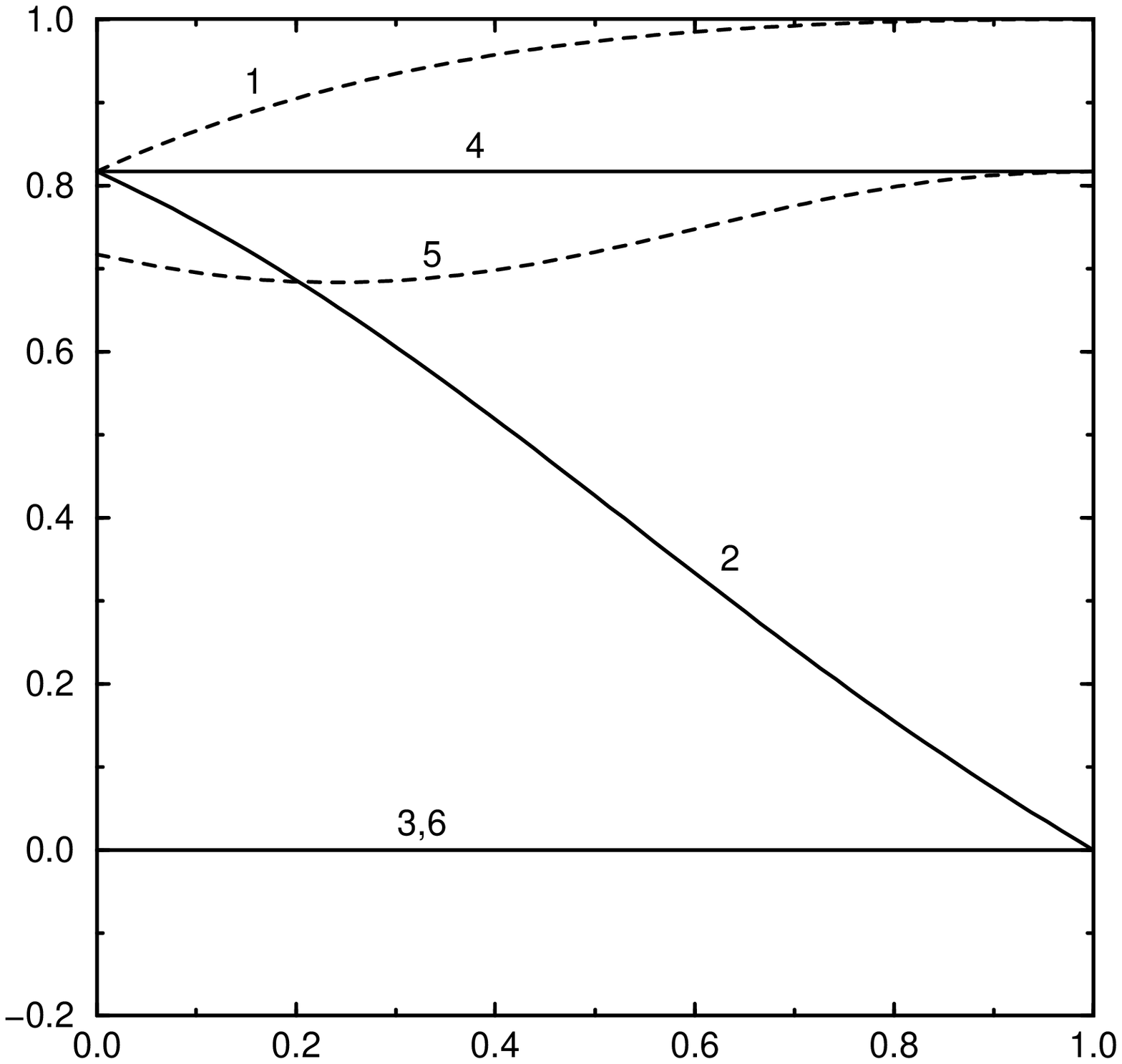,height=14cm,width=13cm} } \\
  \[\;\;\;\;\;\;\;\;\;\;\;\;c_x/c_y \]

\newpage
\thispagestyle{empty}
  \parbox{1.5cm}{\vfill $$ \left<e^3\right> $$ \vfill }
  \parbox{13.5cm}{\epsfig{file=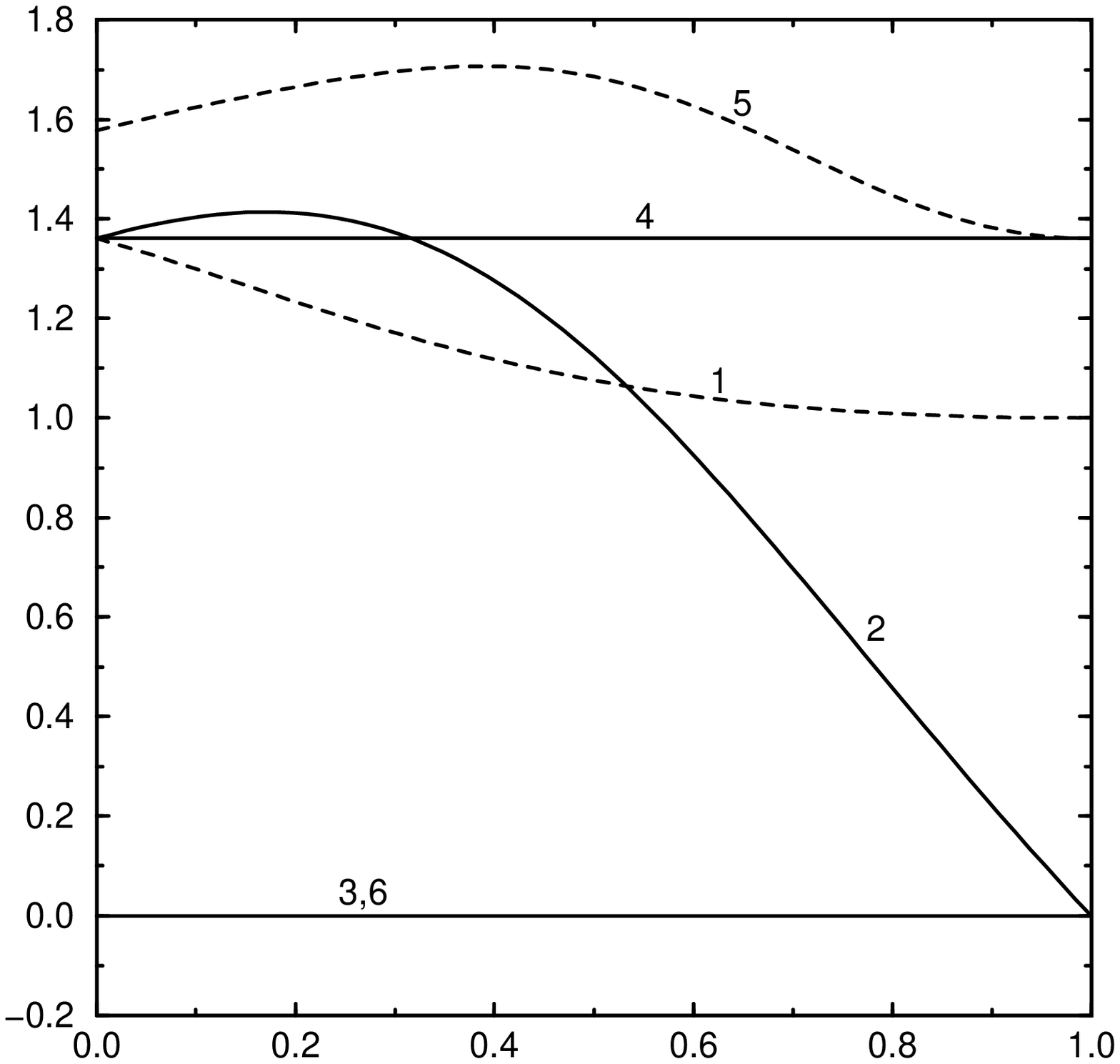,height=14cm,width=13cm} } \\
  \[\;\;\;\;\;\;\;\;\;\;\;\;c_x/c_y \]

\newpage
\thispagestyle{empty}
  \parbox{1.5cm}{\vfill $$ \left<e^4\right> $$ \vfill }
  \parbox{13.5cm}{\epsfig{file=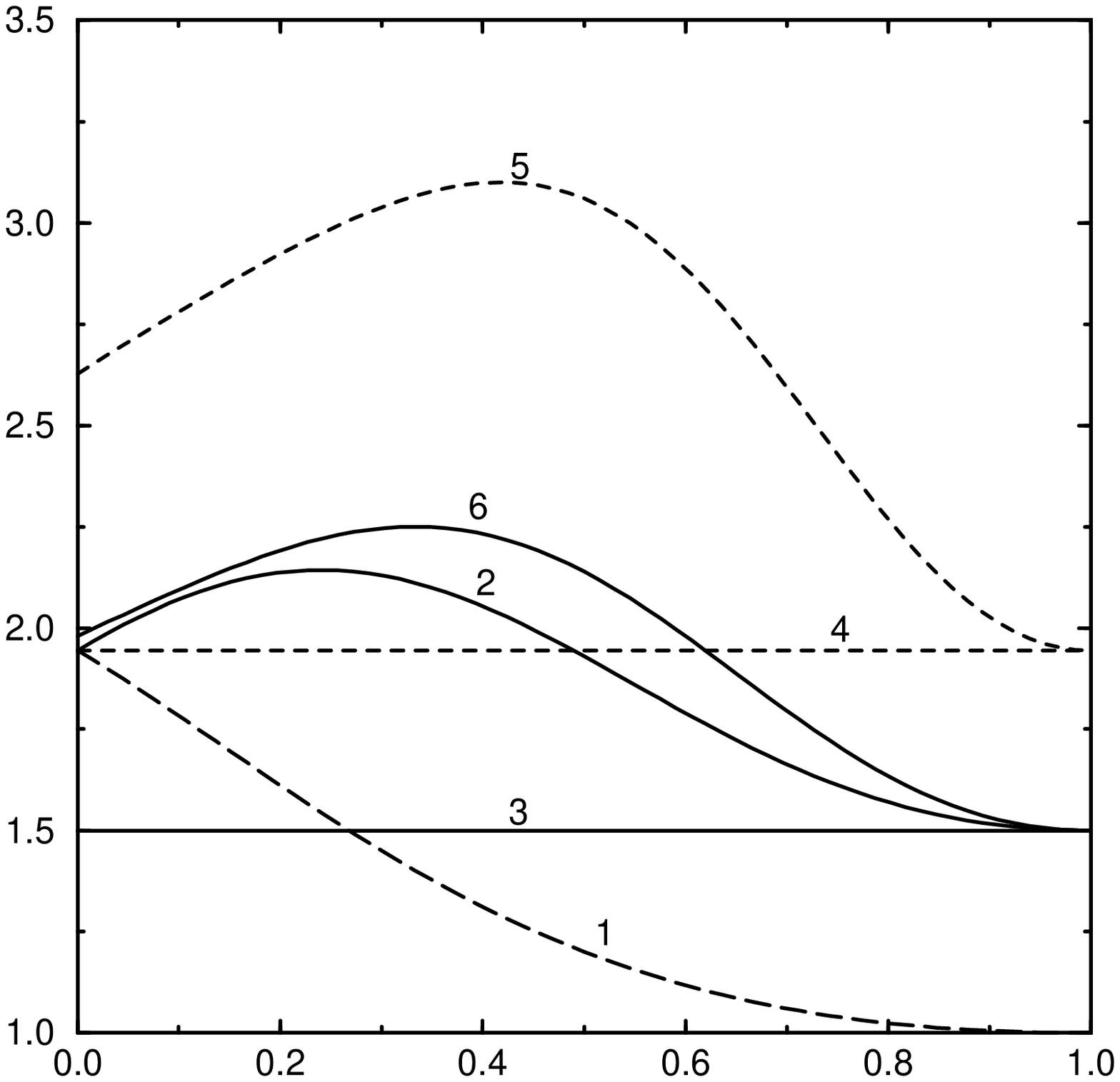,height=14cm,width=13cm} } \\
  \[\;\;\;\;\;\;\;\;\;\;\;\;c_x/c_y \]

\newpage 
\thispagestyle{empty}
  \parbox{1.5cm}{\vfill $$ \frac{T_c}{T_{c_{0}}} $$ \vfill }
  \parbox{13.5cm}{\epsfig{file=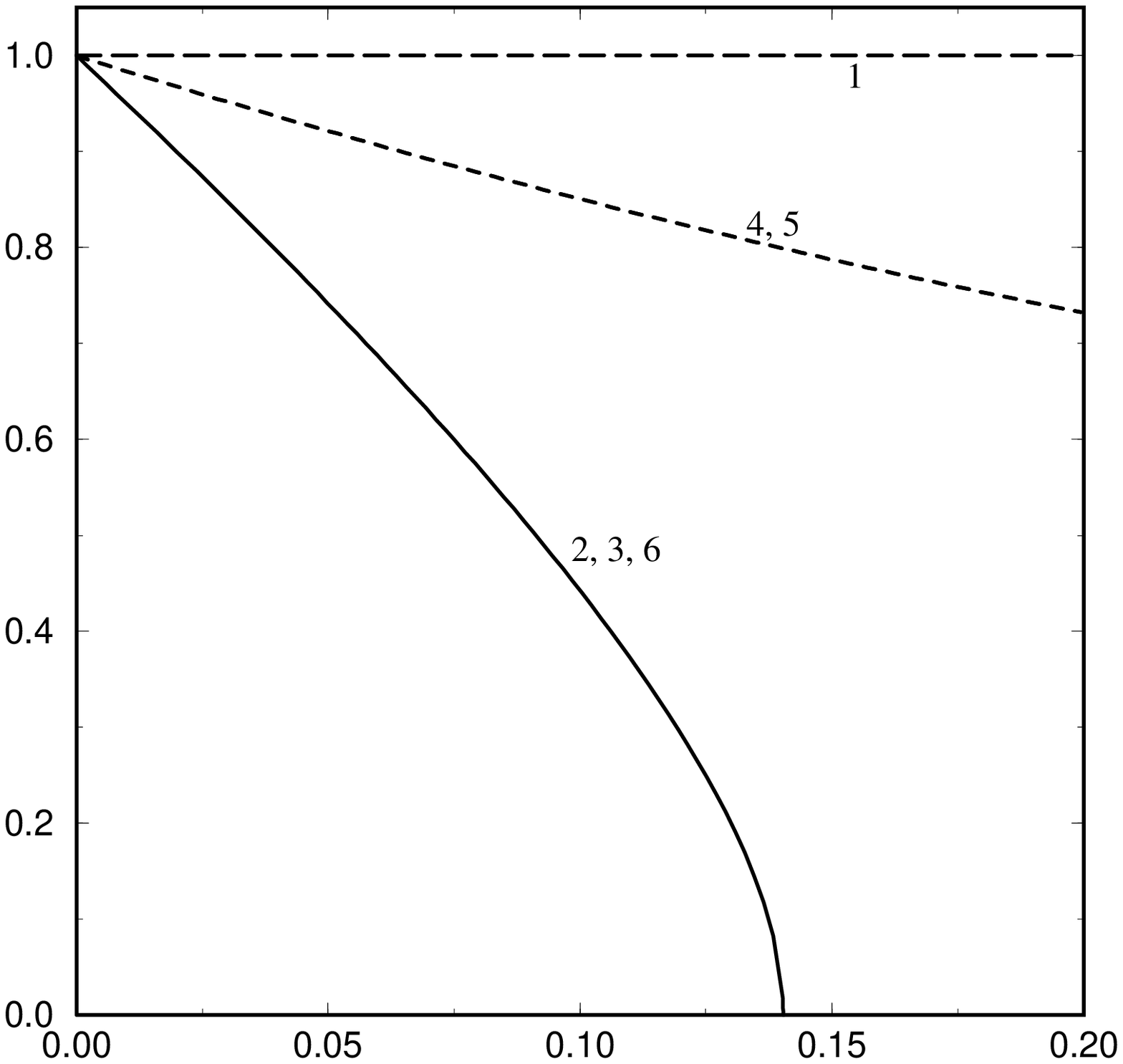,height=14cm,width=13cm} } \\
  \[\;\;\;\;\;\;\;\;\;\;\;\;\varrho_c T_c/T_{c_{0}} \]

\newpage
\thispagestyle{empty}
  \parbox{1.5cm}{\vfill $$ \frac{T_c}{T_{c_{0}}} $$ \vfill }
  \parbox{13.5cm}{\epsfig{file=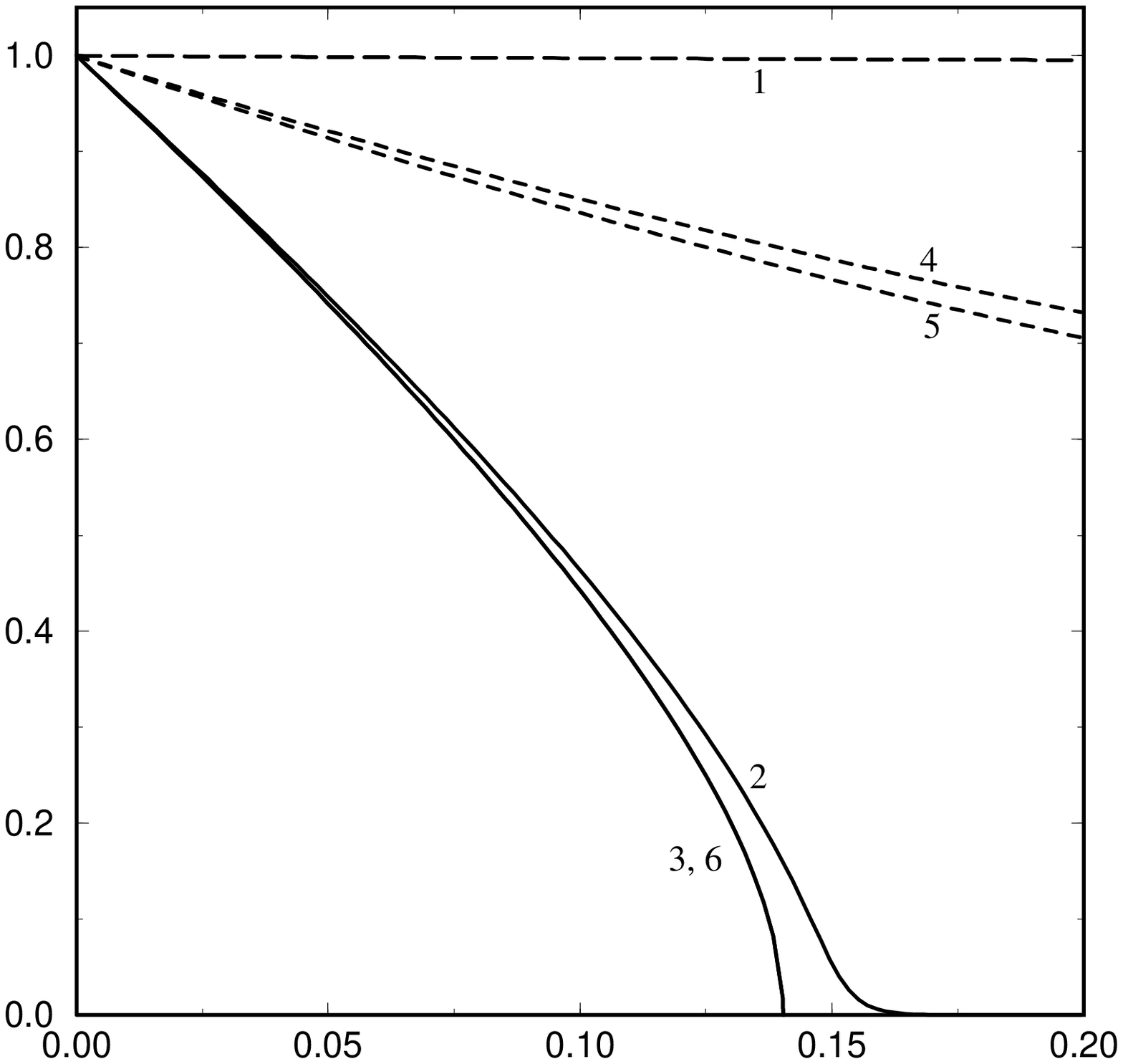,height=14cm,width=13cm} } \\
  \[\;\;\;\;\;\;\;\;\;\;\;\;\varrho_c T_c/T_{c_{0}} \]

\newpage
\thispagestyle{empty}
  \parbox{1.5cm}{\vfill $$ \frac{T_c}{T_{c_{0}}} $$ \vfill }
  \parbox{13.5cm}{\epsfig{file=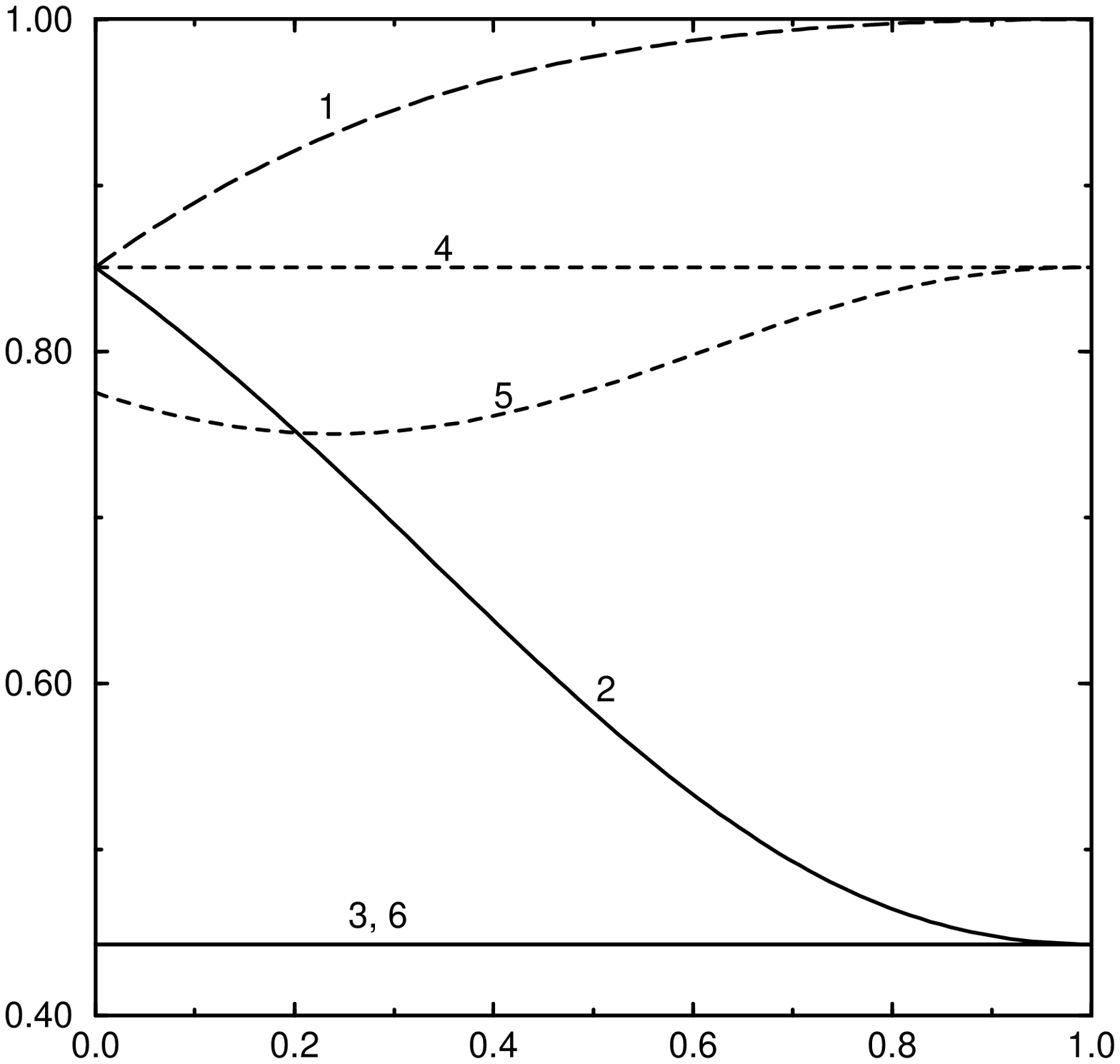,height=14cm,width=13cm} } \\
  \[\;\;\;\;\;\;\;\;\;\;\;\;c_x/c_y \]

\newpage
\thispagestyle{empty}
  \parbox{1.5cm}{\vfill $$ \frac{\Delta C(T_c)}{C_N(T_c)} $$ \vfill }
  \parbox{13.5cm}{\epsfig{file=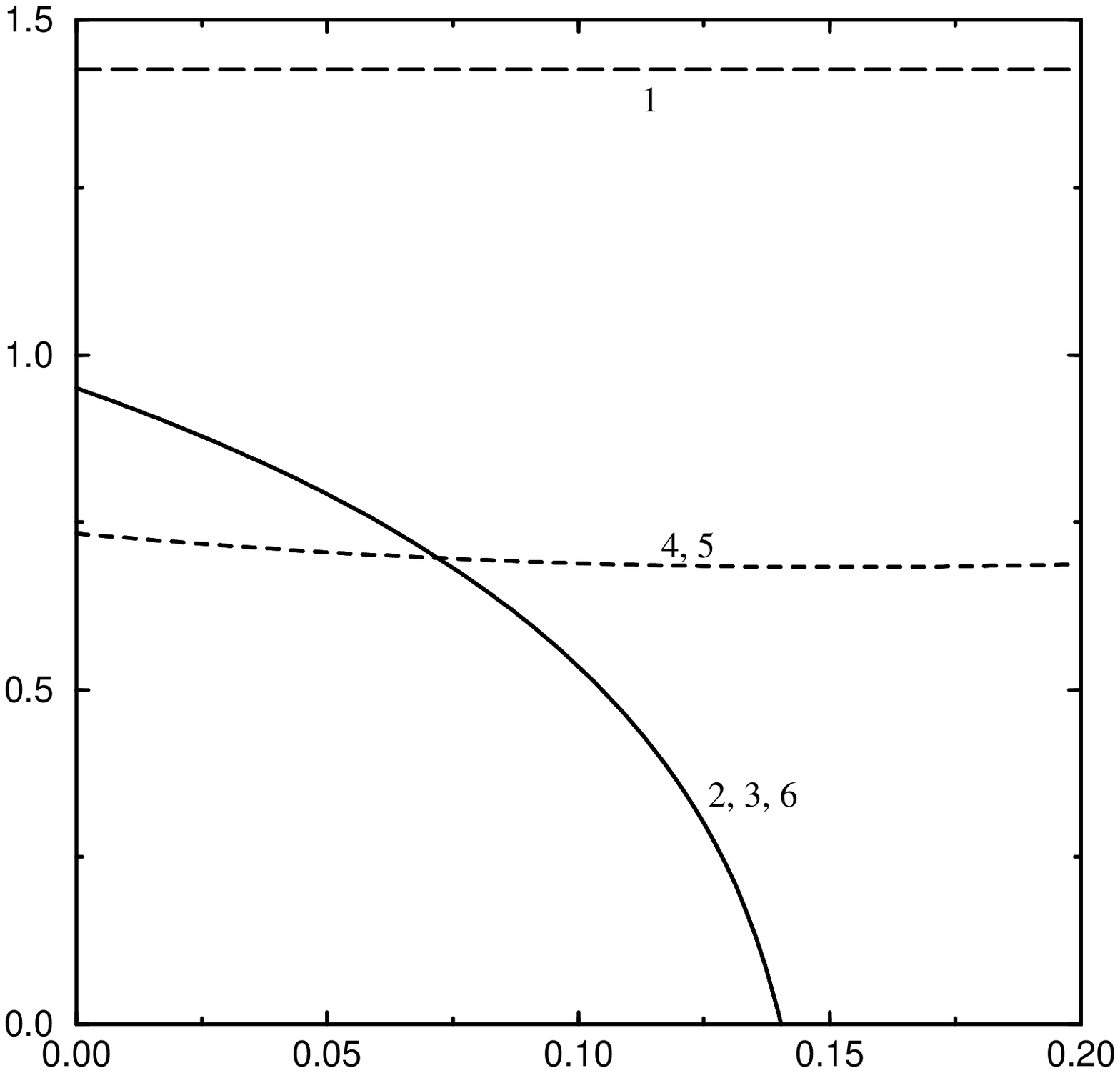,height=14cm,width=13cm} } \\
  \[\;\;\;\;\;\;\;\;\;\;\;\;\varrho_c T_c/T_{c_{0}} \]

\newpage
\thispagestyle{empty}
  \parbox{1.5cm}{\vfill $$ \frac{\Delta C(T_c)}{C_N(T_c)} $$ \vfill }
  \parbox{13.5cm}{\epsfig{file=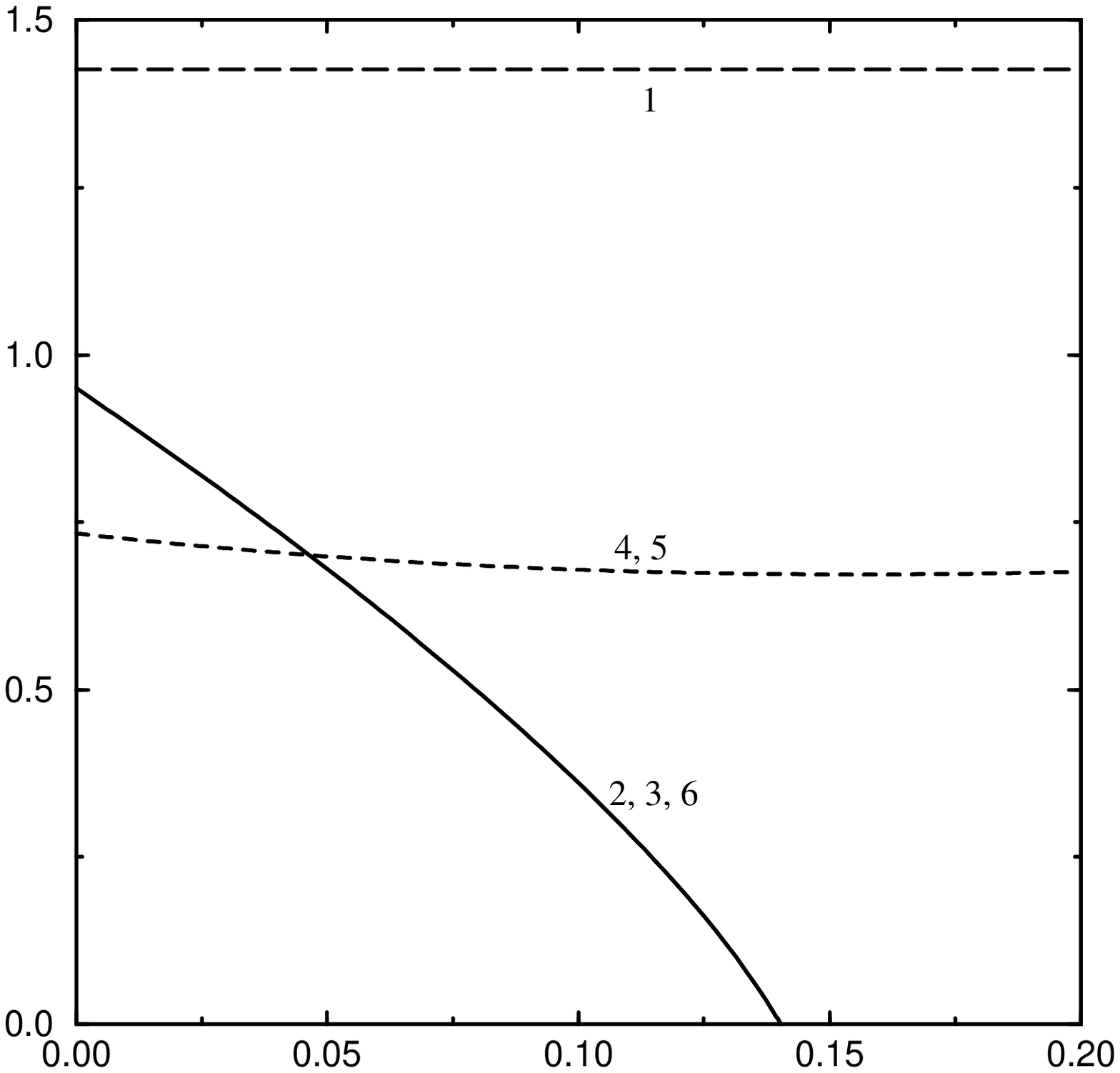,height=14cm,width=13cm} } \\
  \[\;\;\;\;\;\;\;\;\;\;\;\;\varrho_c T_c/T_{c_{0}} \]

\newpage
\thispagestyle{empty}
  \parbox{1.5cm}{\vfill $$ \frac{\Delta C(T_c)}{C_N(T_c)} $$ \vfill }
  \parbox{13.5cm}{\epsfig{file=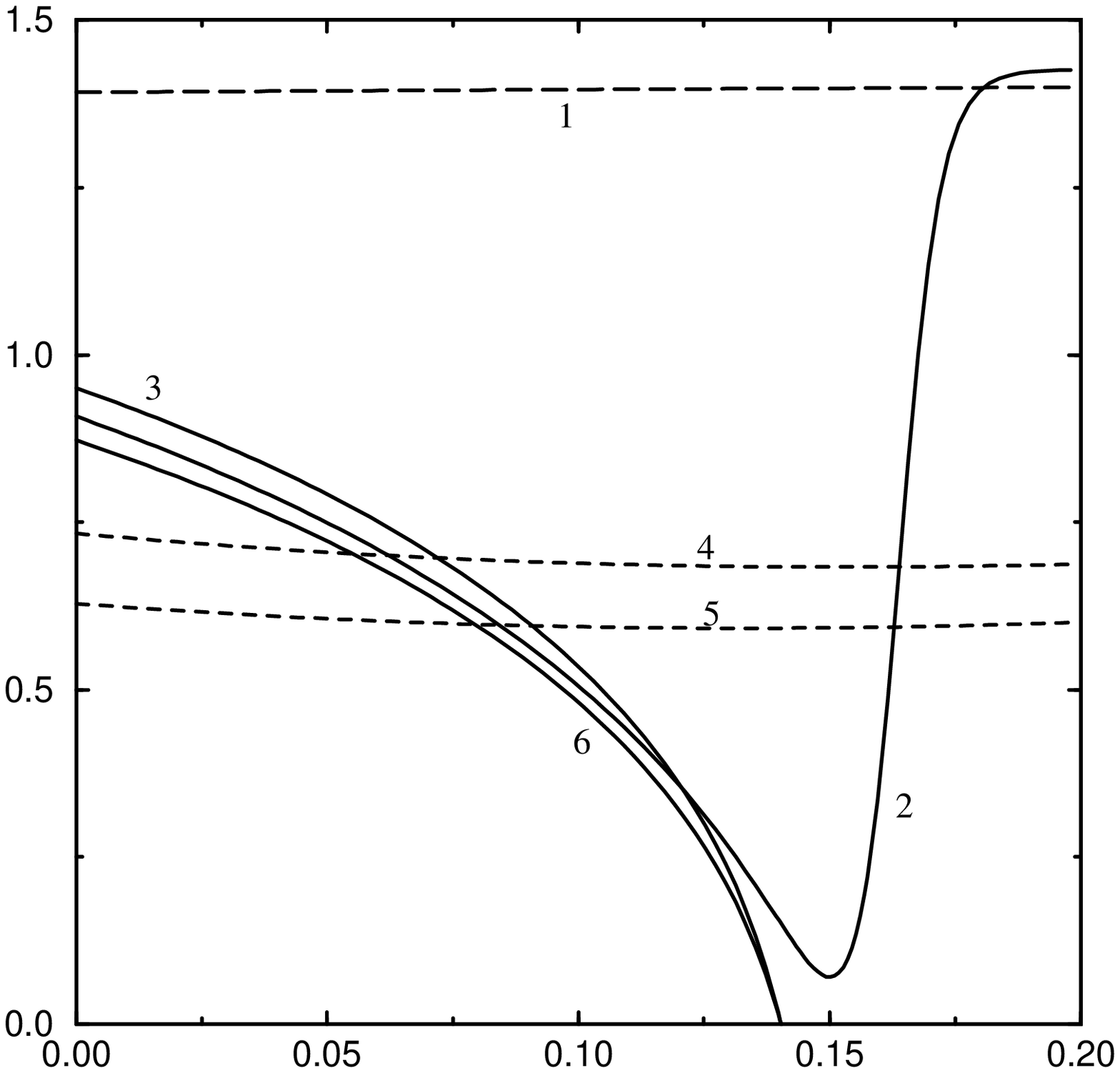,height=14cm,width=13cm} } \\
  \[\;\;\;\;\;\;\;\;\;\;\;\;\varrho_c T_c/T_{c_{0}} \]

\newpage
\thispagestyle{empty}
  \parbox{1.5cm}{\vfill $$ \frac{\Delta C(T_c)}{C_N(T_c)} $$ \vfill }
  \parbox{13.5cm}{\epsfig{file=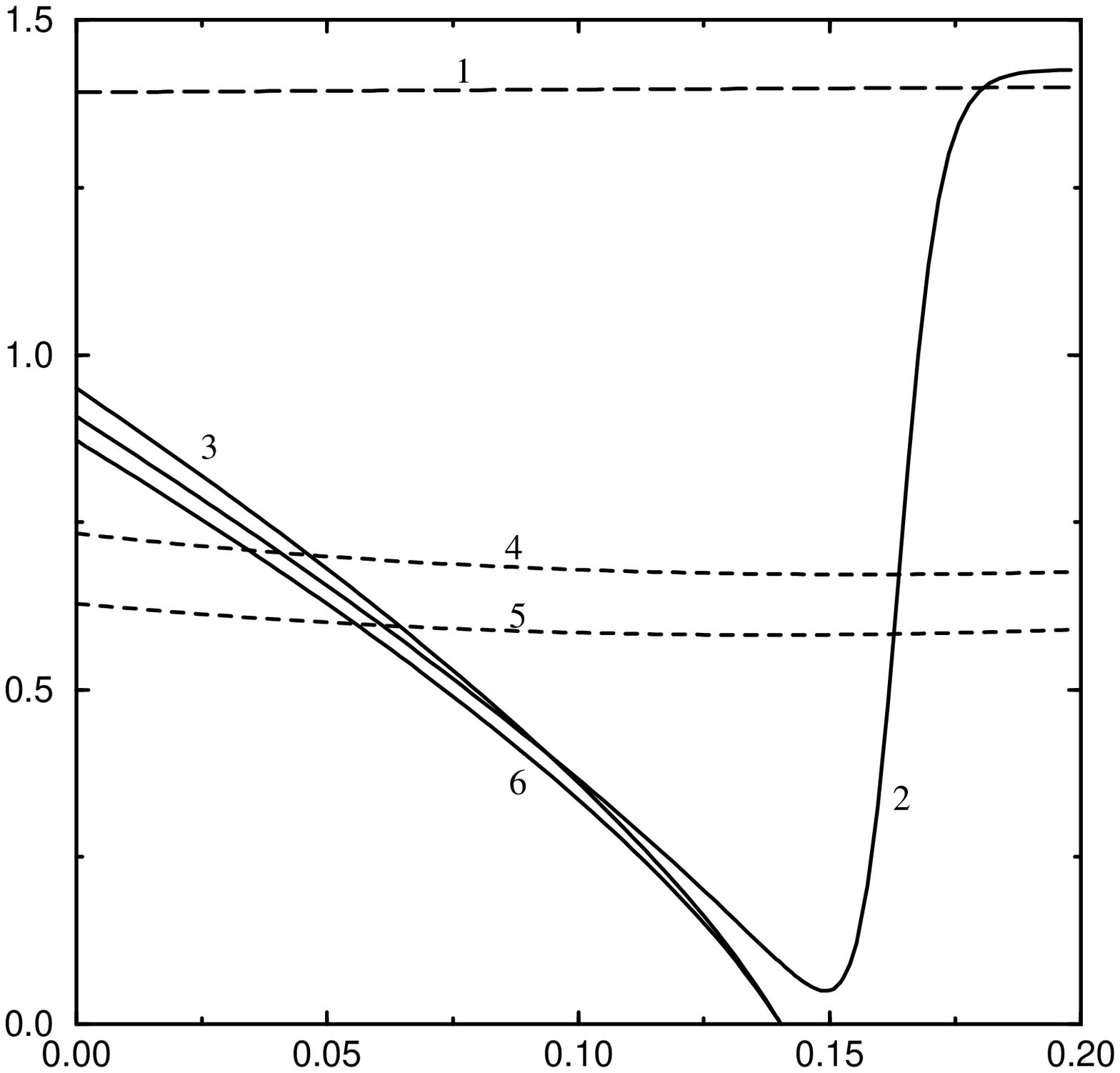,height=14cm,width=13cm} } \\
  \[\;\;\;\;\;\;\;\;\;\;\;\;\varrho_c T_c/T_{c_{0}} \]

\newpage
\thispagestyle{empty}
  \parbox{1.5cm}{\vfill $$ \frac{\Delta C(T_c)}{C_N(T_c)} $$ \vfill }
  \parbox{13.5cm}{\epsfig{file=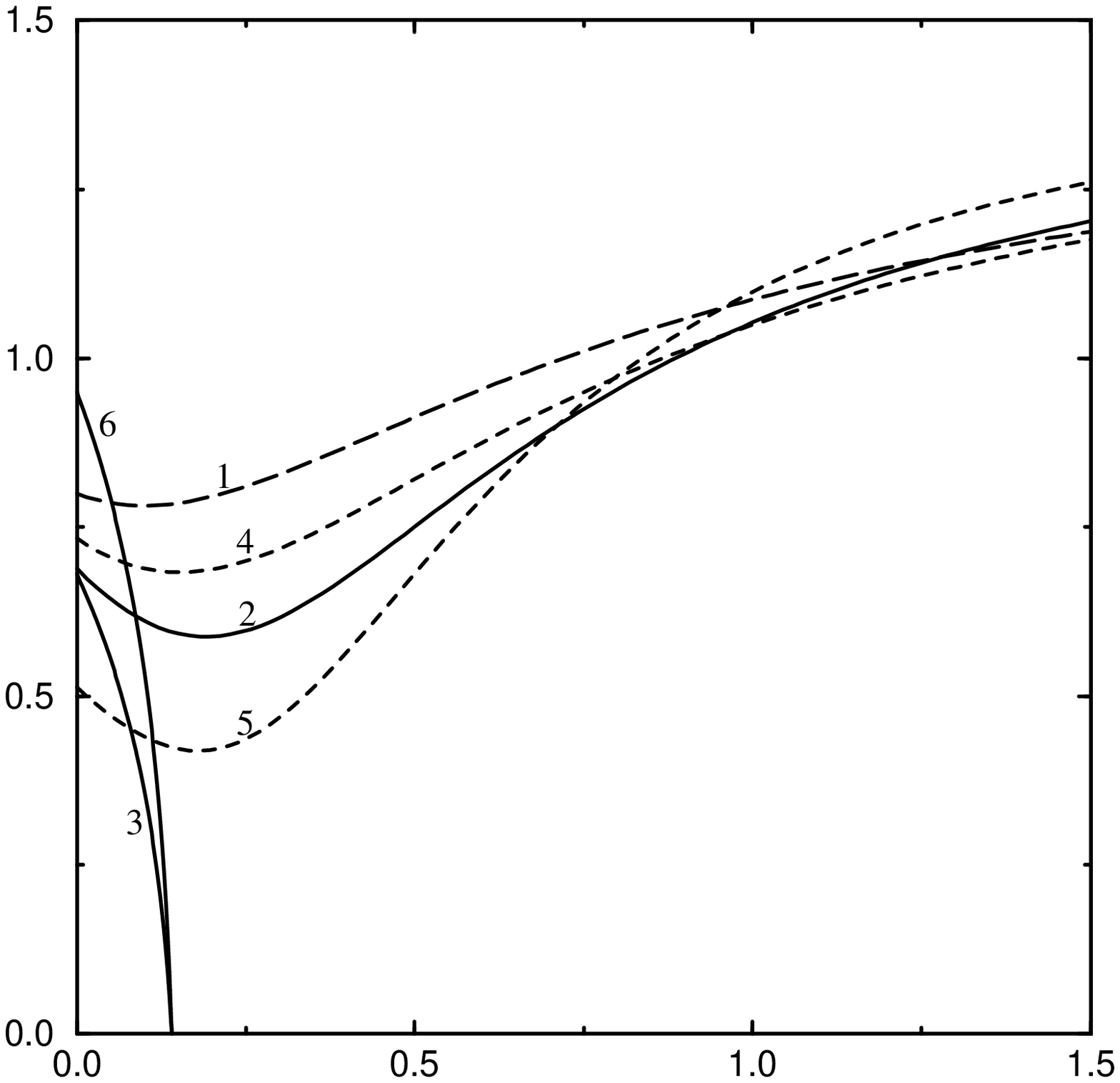,height=14cm,width=13cm} } \\
  \[\;\;\;\;\;\;\;\;\;\;\;\;\varrho_c T_c/T_{c_{0}} \]

\newpage
\thispagestyle{empty}
  \parbox{1.5cm}{\vfill $$ \frac{\Delta C(T_c)}{C_N(T_c)} $$ \vfill }
  \parbox{13.5cm}{\epsfig{file=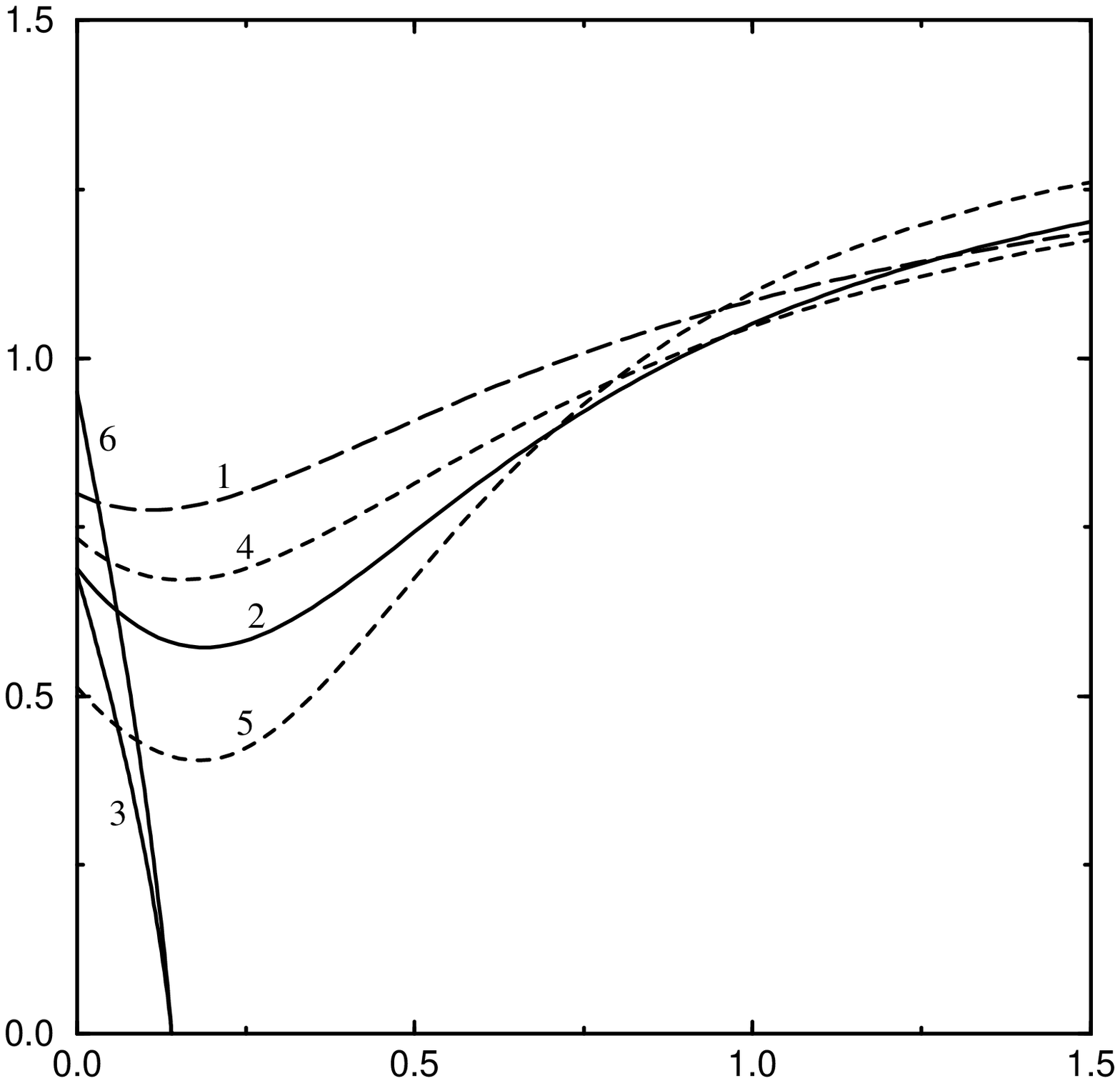,height=14cm,width=13cm} } \\
  \[\;\;\;\;\;\;\;\;\;\;\;\;\varrho_c T_c/T_{c_{0}} \]


\newpage
\begin{references}
\bibitem[*]{AA} on leave from: Institute of Physics, Politechnika
Wroc{\l}awska, Wybrze\.ze Wyspia\'nskiego 27, 50-370 Wroc{\l}aw,
Poland
\bibitem{1} S. A. Sunshine et al., Phys. Rev. {\bf B38}, 893 
(1988); P. Bordet et al., Physica C {\bf 156}, 189 (1988); 
V. Petricek et al.,  Phys. Rev. {\bf B42}, 387 (1990) 
\bibitem{2} D. N. Basov, R. Liang, D. A. Bonn, W. N. Hardy, B. Dabrowski, 
M. Quijada, D. B. Tanner, J. P. Rice, D. M. Ginsberg and T. Timusk, 
Phys. Rev. Lett. {\bf 74}, 598 (1995) 
\bibitem{3} K. Zhang, D. A. Bonn, S. Kamal, R. Liang, D. J. Baar, 
W. N. Hardy, D. N. Basov and T. Timusk, Phys. Rev. Lett. {\bf 73}, 
2484 (1994) 
\bibitem{5} J. L. Cohn, E. F. Skelton, S. A. Wolf, J. Z. Liu and 
R. N. Shelton, Phys. Rev. {\bf B45}, 13144 (1992)
\bibitem{6} R. C. Yu, M. B. Salamon, J. P. Lu and W. C. Lee, 
Phys. Rev. Lett. {\bf 69}, 1431 (1992) 
\bibitem{7} Z. Schlesinger, R. T. Collins, L. D. Rotter, F. Holtzberg, 
C. Feild, U. Welp, G. W. Crabtree, J. Z. Liu, Y. Fang, K. G. Vandervoort  
and S. Fleshler, Physica C {\bf 235-240}, 49 (1994)
\bibitem{4} T. A. Friedmann, M. W. Rabin, J. Giapintzakis, J. P. Rice 
and D. M. Ginsberg, Phys. Rev. {\bf B42}, 6217 (1990) 
\bibitem{11} R. Gagnon, C. Lupien and L. Taillefer,  Phys. Rev. {\bf B50}, 
3458 (1994) 
\bibitem{8} M. A. Quijada, D. B. Tanner, R. J. Kelley and M. Onellion, 
 Physica C {\bf 235-240}, 1123 (1994) 
\bibitem{9} P. B. Allen, W. E. Pickett and H. Krakauer, Phys. Rev. {\bf B37}, 
7482 (1988) 
\bibitem{10} C. O'Donovan and J. P. Carbotte, Phys. Rev. {\bf B52}, 4568 
(1995) 
\bibitem{12} J. Annett, N. Goldenfeld and A. J. Leggett,
in {\it Physical Properties of High Temperature Superconductors},
Vol. 5, D. M. Ginsberg (ed.), (World Scientific, Singapore, 1996)
\bibitem{21} G. Hara\'n, J. Taylor and A. D. S. Nagi, preprint 
\bibitem{13} G.F. Koster et al, {\it Properties of the thirty-two 
point groups} (M.I.T. Press, Cambridge, 1963)
\bibitem{14} H. Kim and E. J. Nicol, Phys. Rev. {\bf B52}, 13576 (1995)
\bibitem{15} P. J. Hirschfeld, D. Vollhardt and P. W\"olfle, Solid State 
Commun. {\bf 59}, 111 (1986); J. Keller, K. Scharnberg and H. Monien, 
Physica C {\bf 152}, 302 (1988); C. J. Pethick and D. Pines, Phys. Rev. Lett. 
{\bf 57}, 118 (1986)  
\bibitem{16} P. J. Hirschfeld, P. W\"olfle and D. Einzel, Phys. Rev. 
{\bf B37}, 83 (1988) 
\bibitem{27} The role of momentum-dependent impurity potential in anisotropic 
superconductors is discussed in  
G. Hara\'n and A. D. S. Nagi, Phys. Rev. {\bf B54} (1996) 
\bibitem{17} K. Maki, in {\it Superconductivity}, R. D. Parks (ed.),
(Marcel Dekker, New York, 1969), Vol. 2, pp. 1035-1102
\bibitem{19} P. J. Davis, in {\it Handbook of Mathematical Functions},
M. Abramowitz and I. A. Stegun (eds.) (Dover, New York, 1965)
pp 253-266
\bibitem{28} Since the normalization $\left<e^2\right>=1$ we do not consider 
the $\left<e^2\right>$ average. One can relax that constraint  
by replacing $e\left({\bf k}\right)$ with $e\left({\bf k}\right)/
\left<e^2\right>^{1/2}$ in Eqs. (17) and (18).   
\bibitem{29} In the BCS theory $T_{c_{0}}=\left(2\omega_D e^{\gamma}/\pi\right)
exp\left\{-1/\left(N_0 V_0 \left<e^2\right>\right)\right\}$, where 
$\omega_D$ is a cut-off energy and $\gamma$ is the Euler constant. 
\bibitem{22} A. A. Abrikosov, Physica C {\bf 214}, 107 (1993)
\bibitem{23} R. J. Radtke, K. Levin, H.-B. Sch\"uttler and M. R. Norman, 
Phys. Rev. {\bf B48}, 653 (1993)
\bibitem{26} Note that the normalized scattering
rate $\varrho_c T_c/T_{c_{0}}$ has a different meaning in the  Born limit,   
where $\varrho_c T_c/T_{c_{0}}=\pi N_0 n_i V^2_i/(2\pi T_{c_{0}})$ and in the 
unitary limit with $\varrho_c T_c/T_{c_{0}}=\Gamma/(2\pi T_{c_{0}})$.
\bibitem{20} Y. Suzumura and H. J. Schulz, Phys. Rev. {\bf B39}, 
11398 (1989) 
\bibitem{24} L. S. Borkowski and P. J. Hirschfeld, Phys. Rev. {\bf B49}, 
15404 (1994)
\bibitem{25} R. Fehrenbacher and M. R. Norman, Phys. Rev. {\bf B50}, 
3495 (1994) 
\end{references}
\end{document}